# Resource Allocation for Energy-Efficient Device-to-Device Communication in 4G Networks

Sajjad Mehri Alamouti and Ahmad R. Sharafat
Department of Electrical and Computer Engineering, Tarbiat Modares University, P. O. Box 14115-194, Tehran, Iran
Email: {sajjad.alamouti, ahmad.sharafat}@gmail.com

*Abstract*— Device-to-device (D2D) communications as an underlay of a LTE-A (4G) network can reduce the traffic load as well as power consumption in cellular networks by way of utilizing peer-to-peer links for users in proximity of each other. This would enable other cellular users to increment their traffic, and the aggregate traffic for all users can be significantly increased without requiring additional spectrum. However, D2D communications may increase interference to cellular users (CUs) and force CUs to increase their transmit power levels in order to maintain their required quality-of-service (QoS). This paper proposes an energy-efficient resource allocation scheme for D2D communications as an underlay of a fully loaded LTE-A (4G) cellular network. Simulations show that the proposed scheme allocates cellular uplink resources (transmit power and channel) to D2D pairs while maintaining the required QoS for D2D and cellular users and minimizing the total uplink transmit power for all users.

## I. Introduction

The broadband cellular network is a technology enabler and an indispensable driving force for further economic growth and development [1]. It is predicted that by the end of 2014, the number of mobile-connected devices (including machine-to-machine modules) will exceed the number of people on the earth, and by 2018 there will be more than 10 billion mobile-connected devices in the world [2]. However, the frequency spectrum needed for broadband cellular networks in existing frameworks is not sufficient, and there is a need to introduce novel schemes for more efficient use of available resources. Device-to-device (D2D) communications as an underlay of a LTE-A cellular network can alleviate the need for more resources to some extent while increasing the users' aggregate capacity [3].

In cellular networks, adjacent users may be able to set up a direct D2D link using the cellular interface, and subsequently exchange data via the D2D link without traversing a base station (BS) or the core network [4]. In doing so, the transmitter of a D2D pair (D_Tx) utilizes the spectrum of the cellular network to transmit to its receiver (D_Rx) via the D2D link. Since cellular users (CUs) and D2D pairs simultaneously use the same spectrum, they may interfere with each other, and hence, there is a need for interference management [5]. Channel (spectrum) allocation and transmit power control schemes are the two widely used interference management techniques in wireless networks [6]-[9].

Furthermore, energy efficiency in cellular networks is a growing concern [10] and D2D communication can reduce power consumption in base stations and cellular users [11]. However, when a D2D pair occupy an uplink channel of a CU, the CU will increase its transmit power to maintain its QoS. In [9], an uplink resource allocation scheme for D2D links as an underlay of a fully loaded cellular network is proposed that maximizes the overall network throughput while maintaining the QoS for both D2D pairs and CUs. In this scheme, when a D2D pair utilize a CU's uplink spectrum, at least one user (the D_Tx or the CU) transmits at its peak power to maximize the overall throughput. For a system model similar to that in [9], a distributed power control scheme for D2D users is proposed in [7], where such users can reuse the CUs' uplink channels in an opportunistic manner only when their interference to the base station is less than the margin $k$ in the latter's required SINR. In practice, this forces the CUs to increase their transmit power levels to maintain their SINRs and their throughputs, which may not be practical in power constrained CUs or may reduce energy efficiency. Hence, it is desirable to devise a new scheme in which the throughput is increased to the extent possible while satisfying the QoS for all users and the increase in the uplink transmit power is minimized.

In this paper, we propose a new resource allocation scheme for D2D links as an underlay of a fully loaded cellular LTE-A network. In our proposed scheme, the corresponding base station optimally allocates the uplink resources for each D2D link by minimizing the total transmit power levels for D2D pairs and CUs while satisfying the required QoS for both user types. We assume orthogonal frequency-division multiple access (OFDMA) for the downlink and single-carrier FDMA (SC-FDMA) for the uplink of each cellular user [12], where the uplink data spreads across multiple sub-carriers. In doing so, similar to [9], the base station first checks if admitting a D2D pair as an underlay of the cellular network would not violate the QoS requirements for both the D2D pair and its potential CU partners, and subsequently determines their minimum required transmit power levels. Then, a matching CU partner is identified for which the overall power consumption in the network is minimized.

The rest of this paper is organized as following. The system model with integrated D2D links is given in Section II, followed by formulation of the optimization problem for minimizing the total uplink transmit power in Section III. The optimal resource allocation algorithm is proposed and analyzed in Section IV. Simulation results are in Section V, and conclusions are in Section VI.

This work was supported in part by Tarbiat Modares University, Tehran, Iran.

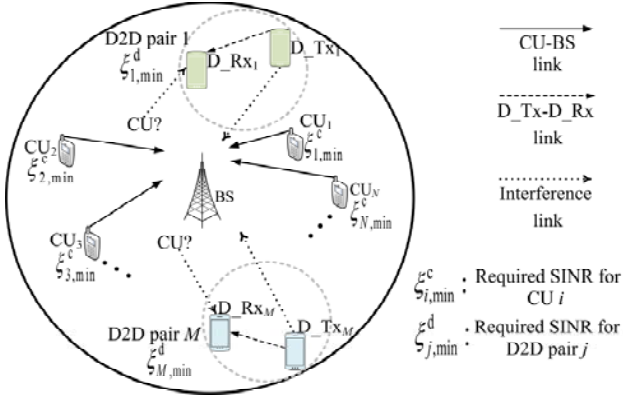

Fig. 1. System model of D2D communications as an underlay of a cellular network where D2D pairs and CUs share the same uplink channels [9].

## II. System Model

### A. Infrastructure and User Model

Similar to [9], we assume a fully loaded cellular network in which all channels are assigned to CUs and there is no spare channel for D2D pairs. As in Fig. 1, we consider a single cell with radius $R$ in which $N$ active CUs occupy $N$ orthogonal uplink channels, and there is no spare uplink channel for $M$ D2D pairs that are situated in that cell. The base station is equipped with omni-directional antennas at the center of the cell, and has perfect CSI of all links in the cell. The required QoS in terms of required SINR for CU $i$ is $\xi^c_{i,\min}$ and for D2D pair $j$ is $\xi^d_{j,\min}$. The set of active CUs is $\mathcal{C} = \{1,...,N\}$ and the set of D2D pairs is $\mathcal{D} = \{1,...,M\}$. Each D2D link is considered as an underlay of the cellular network. This means that D2D transmissions and cellular uplink transmissions may interfere with each other.

### B. Channel Model

The channel model is as in [9]. In addition to the distanced-based path loss, both fast fading due to multi-path propagation and slow fading due to shadowing are considered. Hence, the uplink channel gain between CU $i$ and the base station is

$$g_{i,B} = K\beta_{i,B}\zeta_{i,B}L_{i,B}^{-\alpha}, \quad (1)$$

where $K$ is a constant that depends on system parameters, $\beta_{i,B}$ is the fast fading gain with exponential distribution, $\zeta_{i,B}$ is the slow fading gain with log-normal distribution, $\alpha$ is the path loss exponent, and $L_{i,B}$ is the distance between CU $i$ and the base station. Similarly, the channel gain between D2D pair $j$ is

$$g_j = K\beta_j\zeta_j L_j^{-\alpha}, \quad (2)$$

and channel gains of the interference links from the transmitter of the D2D pair $j$ (D_Tx$_j$) to the base station, denoted by $h_{j,B}$, and from the CU $i$ to the receiver of the D2D pair $j$ (D_Rx$_j$), denoted by $h_{i,j}$ are

$$h_{j,B} = K\beta_{j,B}\zeta_{j,B}L_{j,B}^{-\alpha}, \quad (3)$$

$$h_{i,j} = K\beta_{i,j}\zeta_{i,j}L_{i,j}^{-\alpha}, \quad (4)$$

respectively. We assume additive white Gaussian noise with power $\sigma_N^2$ in each channel.

## III. Resource Allocation Problem Formulation

A D2D link can only be set up when the required SINR values for both the D2D pair and CUs can be guaranteed with bounded transmit power levels. When this criterion is met, the D2D pair is considered admissible, and the corresponding CU that occupy the same channel is called its reuse partner. We wish to maximize the number of admissible D2D pairs and minimize the total uplink transmit power of CUs and admissible D2D pairs. Let $P_i^c$ and $P_j^d$ denote the transmit power of CU $i$ and the transmitter of D2D pair $j$, $\xi_i^c$ and $\xi_j^d$ denote the SINR of CU $i$ and D2D pair $j$, and $P_{\max}^c$ and $P_{\max}^d$ denote the maximum transmit power levels of CUs and transmitters of D2D pairs, respectively. We consider a scenario where at most one CU and one D2D pair share the same uplink channel. In this case, the value of the integer variable $\rho_{i,j} \in \{0,1\}$ is 1, otherwise $\rho_{i,j} = 0$, and an assignment matrix $\rho = [\rho_{i,j}]_{N \times M}$ is formed. The optimization problem is mathematically formulated as

$$\text{Determine} \begin{cases} \rho = [\rho_{i,j}]_{N \times M}, \\ P_i^c, \forall i \in \mathcal{C}, \\ P_j^d, \forall j \in \mathcal{D}, \end{cases} \quad (5)$$

$$\text{To Maximize} \sum_{i=1}^N \sum_{j=1}^M \rho_{i,j}, \quad (5a)$$

$$\text{To Minimize} \sum_{i=1}^N \sum_{j=1}^M (P_i^c + \rho_{i,j}P_j^d), \quad (5b)$$

Subject to:
$$\begin{cases} \xi_i^c = \dfrac{P_i^c g_{i,B}}{\sigma_N^2 + \rho_{i,j}P_j^d h_{j,B}} \geq \xi_{i,\min}^c, \forall i \in \mathcal{C}, & (5c) \\[6pt] \xi_j^d = \dfrac{P_j^d g_j}{\sigma_N^2 + \rho_{i,j}P_i^c h_{i,j}} \geq \xi_{j,\min}^d, \forall j \in \mathcal{S}, & (5d) \\[6pt] \sum_{j=1}^M \rho_{i,j} \leq 1, \ \rho_{i,j} \in \{0,1\}, \forall i \in \mathcal{C}, & (5e) \\[6pt] \sum_{i=1}^N \rho_{i,j} \leq 1, \ \rho_{i,j} \in \{0,1\}, \forall j \in \mathcal{D}, & (5f) \\[6pt] 0 < P_i^c \leq P_{\max}^c, \forall i \in \mathcal{C}, & (5g) \\[6pt] 0 \leq P_j^d \leq P_{\max}^d, \forall j \in \mathcal{D}, & (5h) \end{cases}$$

where $\mathcal{S}$ ($\mathcal{S} \subseteq \mathcal{D}$) is the set of admissible D2D pairs. Constraints (5c) and (5d) represent the QoS requirements of CUs and admissible D2D pairs in terms of their SINRs, respectively. Constraint (5e) ensures that at most one D2D pair can reuses the uplink channel of a CU, while constraint (5f) ensures that a D2D pair can reuses at most one CU's uplink channel. Constraints (5g) and (5h) indicate that transmit power levels are bounded.

The optimization problem in (5) is a <u>m</u>ixed <u>i</u>nteger <u>l</u>inear <u>p</u>rogramming (MILP) problem, and is difficult to solve in a direct manner. In the following section, we will divide the problem in (5) into two sub problems, and solve them individually.

## IV. Optimal Resource Allocation

We will solve the optimization problem in (5) by dividing it into two sub problems. The first one is to check for admissibility of each D2D pair and to obtain the minimal transmit power levels for all admissible D2D pairs and their candidate reuse partners that can satisfy their respective SINR requirements. The second one is to identify the optimal uplink channel for all admissible D2D pairs among the uplink channels of all candidate reuse partners.

### A. Admissibility and Optimal Power Control of D2D Pairs

We utilize the scheme proposed in [9] for admissibility and determining candidate CU reuse partners, and for brevity, do not repeat it here. From [9], the minimum transmit power levels for CU $i$ as a candidate reuse partner and for D2D pair $j$ that satisfy their required SINRs are

$$\begin{cases} P_{i,A}^c = \dfrac{(g_j \xi_{i,\min}^c + h_{j,B} \xi_{i,\min}^c \xi_{j,\min}^d) \sigma_N^2}{g_j g_{i,B} - \xi_{i,\min}^c \xi_{j,\min}^d h_{i,j} h_{j,B}}, \\ P_{j,A}^d = \dfrac{(h_{i,j} \xi_{i,\min}^c \xi_{j,\min}^d + g_{i,B} \xi_{j,\min}^d) \sigma_N^2}{g_j g_{i,B} - \xi_{i,\min}^c \xi_{j,\min}^d h_{i,j} h_{j,B}}. \end{cases} \quad (6)$$

The D2D pair $j$ is admissible and CU $i$ is its candidate reuse partner if

$$\begin{cases} 0 < P_{i,A}^c \leq P_{\max}^c, \\ 0 < P_{j,A}^d \leq P_{\max}^d. \end{cases} \quad (7)$$

Let $\mathcal{R}_j$ denote the set of candidate reuse partners for the D2D pair $j$. When there is no candidate reuse partner for the D2D pair $j$, we have $\mathcal{R}_j = \varnothing$, $\rho_{i,j} = 0, \forall i \in \mathcal{C}$ and $P_j^d$ is set to zero.

### B. Optimal Resource Allocation for Admissible D2D Pairs

When CU $i$ exclusively uses the uplink channel, its transmit power that satisfies its required SINR is

$$P_{i,\min}^c = \dfrac{\xi_{i,\min}^c \sigma_N^2}{g_{i,B}}. \quad (8)$$

When the D2D pair $j$ reuses the same uplink channel, the transmit power for CU $i$ that satisfies its required SINR increases, i.e., $P_{i,A}^c \geq P_{i,\min}^c$ and their total uplink transmit power is

$$P_{i,j}^{sum} = P_{i,A}^c + P_{j,A}^d. \quad (9)$$

Now, let $P_{i,j}^{inc}$ be the increase in the transmit power levels of CU $i$ and D2D pair $j$ due to their shared use of spectrum, i.e.,

$$P_{i,j}^{inc} = P_{i,j}^{sum} - P_{i,\min}^c. \quad (10)$$

Hence, when there is only one admissible D2D pair $j$ in the cell, one can find its optimal CU reuse partner via

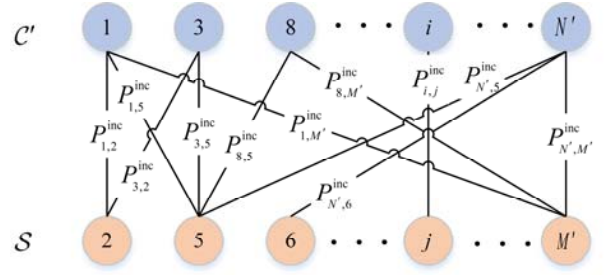

Fig. 2. An example of bipartite graph for the assignment problem.

$$i^* = \arg \min_{i \in \mathcal{R}_j} P_{i,j}^{inc}. \quad (11)$$

When there are multiple admissible D2D pairs in the cell, the problem of finding the optimal reuse partner for each admissible D2D pair is the minimum weighted matching problem on the bipartite graph, or in general, is an assignment problem [13, Ch. 10]. The bipartite graph is also used in [9] but with a different objective. The problem can be formulated as

$$\min_{\rho_{i,j}, P_{i,j}^{inc}} \left\{ \sum_{i \in \mathcal{C}'} \sum_{j \in \mathcal{S}} \rho_{i,j} P_{i,j}^{inc} \right\}, \quad (12)$$

Subject to: $\begin{cases} \sum_{j \in \mathcal{S}} \rho_{i,j} \leq 1, \; \rho_{i,j} \in \{0,1\}, \forall i \in \mathcal{C}', \quad (12a) \\ \sum_{i \in \mathcal{C}'} \rho_{i,j} = 1, \; \rho_{i,j} \in \{0,1\}, \forall j \in \mathcal{S}, \quad (12b) \end{cases}$

where $\mathcal{C}'$ is the union of all candidate reuse partner sets for all D2D pairs, i.e.,

$$\mathcal{C}' = \mathcal{R}_1 \cup \mathcal{R}_2 \cup \ldots \cup \mathcal{R}_M. \quad (13)$$

Fig. 2 illustrates the assignment problem in (12). In this figure, the set of admissible D2D pairs $\mathcal{S}$, and the union of all candidate reuse partner sets for all D2D pairs $\mathcal{C}'$ are the two groups of vertices in the bipartite graph. If CU $i$ is a reuse candidate for D2D pair $j$, vertex $i$ connects to vertex $j$ by an edge $ij$. The increase in transmit power levels of CU $i$ and D2D pair $j$, i.e., $P_{i,j}^{inc}$, is the weight of edge $ij$.

The Hungarian algorithm can be used to solve this problem. This algorithm is an efficient bipartite assignment scheme whose worst case computational complexity is at most $O(n^3)$, where $n$ is the number of vertices in one group of the symmetric bipartite graph [14, Ch. 3].

In general, $|\mathcal{C}'| > |\mathcal{S}|$, but the Hungarian algorithm requires the bipartite graph to be perfectly symmetric. To satisfy this requirement, we add $|\mathcal{C}'| - |\mathcal{S}|$ virtual vertices to the set of admissible D2D pairs $\mathcal{S}$ in the original graph. If the vertex $i$ is not connected to the vertex $j$, we connect them with a large-valued weighted edge. Now, we use the Hungarian algorithm to solve the minimum weighted matching problem on the transformed bipartite graph. The algorithm for solving the optimal resource allocation problem in (5) is presented in Table I.

TABLE I. OPTIMAL RESOURCE ALLOCATION ALGORITHM

| |
|---|
| 1: $\mathcal{C}$ : The set of active CUs in the cell |
| 2: $\mathcal{D}$ : The set of all D2D pairs in the cell |
| 3: $\mathcal{R}_j$ : The set of reuse candidates for D2D pair $j$ |
| 4: $\mathcal{S}$ : The set of all admissible D2D pairs |
| 5: $\rho = [\rho_{i,j}]_{N \times M} = 0$ (assignment initialization) |
| 6: **Step 1** |
| 7: **For all** $j \in \mathcal{D}$ and $i \in \mathcal{C}$ **do** |
| 8:      Calculate $P_{i,A}^c$ and $P_{j,A}^d$ |
| 9:      **If** $0 < P_{i,A}^c \leq P_{max}^c$ & $0 < P_{j,A}^d \leq P_{max}^d$ **then** $i \in \mathcal{R}_j$; $j \in \mathcal{S}$ |
| 10:      **End if** |
| 11: **End for** |
| 12: **Step 2** |
| 13: **If** $|\mathcal{S}| = 1$ **then** |
| 14:      $i^* = \arg \min_{i \in \mathcal{R}_j} P_{i,j}^{inc}$ & $\rho_{i^*,j} = 1$ |
| 15: **Else** |
| 16: Construct the symmetric bipartite graph and use the Hungarian algorithm to obtain $\rho_{i,j}, \forall i \in \mathcal{R}_1 \cup \mathcal{R}_2 \cup \ldots \cup \mathcal{R}_M$ and $\forall j \in \mathcal{S}$ |
| 17: **End if** |

## V. Simulation Results

We consider a single isolated circular cell in which CUs are uniformly distributed and each D2D pair is located in a uniformly distributed cluster with radius $r$. The total uplink bandwidth is equally shared between all active CUs. When no D2D link is active, each CU can reach its required SINR with its constrained transmit power. Parameter values for simulations are summarized in Table II. We compare the performance of our proposed scheme with that in [7] which assumes that there exists a fixed margin $k$ in each CU's required SINR to compensate for the interference caused by D2D transmitters and also assumes that each D2D transmitter, with the knowledge of $k$, adjusts its transmit power to satisfy the CU's QoS requirement. With these assumptions, we utilize the Hungarian algorithm to identify the optimal CU partner for each D2D pair.

We consider four metrics in our simulations. First, in Figs. 3 and 7 we investigate how many D2D pairs are able to reuse CUs' uplink channels in terms of the access ratio, defined as the ratio of the number of admissible D2D pairs to the total number of D2D pairs in the cell. Second, in Figs. 4 and 8 we investigate the increase in the total system uplink power when D2D communications are allowed as compared to when D2D communications are not permitted. Third, in Figs. 5 and 9, we show the increase in the total system uplink throughput when D2D communications are allowed as compared to when D2D communications are not permitted. Finally, in Figs. 6 and 10, we show the increase in the total system uplink energy efficiency defined as the ratio of the total system uplink throughput to the total system uplink power when D2D communications are allowed as compared to when D2D communications are not permitted. Results are the averaged 10,000 realizations.

As expected, simulations show that via our proposed scheme, the underlay D2D communications in cellular networks can increase the users' capacity and the total system uplink throughput with minimal increase in the total system uplink power, and result in an improved energy efficiency.

TABLE II. SIMULATION PARAMETERS

| Parameter | Value |
|---|---|
| Cell radius ($R$) | 0.5, 1 km |
| System uplink bandwidth | 5 MHz |
| AWGN power ($\sigma_N^2$) | -114 dBm |
| Pathloss exponent ($\alpha$) | 4 |
| Pathloss constant ($K$) | $10^{-2}$ |
| Max. CU transmit power ($P_{max}^c$) | 24 dBm |
| Max. D_Tx transmit power ($P_{max}^d$) | 24 dBm |
| Req. SINR for CU $i$ ($\xi_{i,min}^c$) | Uniform distribution in [0,25] dB |
| Req. SINR for D2D pair $j$ ($\xi_{j,min}^d$) | Uniform distribution in [0,25] dB |
| D2D cluster radius ($r$) | 20, 30, 40, …, 100 m |
| Distance between D2D cluster's center and BS | Uniform distribution in [0,$R$] |
| No. of active CUs ($N$) | 20, 40 |
| No. of D2D pairs ($M$) | 10%, 20%, …, 100% of active CUs |
| Fast fading gain ($\beta_{i,B}$) | Exponential distribution with unit mean |
| Slow fading gain ($\zeta_{i,B}$) | Log-normal distribution with unit mean and standard deviation of 8 dB |
| Fixed margin ($k$) | 1dB |

Fig. 3 shows that the access ratio for any given distance between a D2D pair is higher for larger cells, but reduces with increasing the D2D distance. Fig. 4 illustrates that the increase in the total system uplink power for any given D2D distance is less for larger cells, and increases with increasing the D2D distance. Fig. 5 shows that the increase in the total system uplink throughput is higher for larger cells. It also shows that the increase in the total system uplink throughput decreases with increasing the D2D distance. Fig. 6 illustrates that for any given D2D distance, the increase in the total system uplink energy efficiency is higher for larger cells, and decreases with increasing the D2D distance.

Fig. 7 illustrates that the access ratio remains steady irrespective of the ratio of D2D pairs to CUs. Fig. 8 shows that the increase in the total system uplink power is unaffected by increasing the number of CUs, but increases with increasing the number of D2D pairs. Fig. 9 shows that the increase in the total system uplink throughput is unaffected by the increase in the number of CUs, but increases with increasing the number of D2D pairs. Finally, Fig. 10 shows that the increase in the total system uplink energy efficiency is also unaffected by the increase in the number of CUs, but increases with increasing the number D2D pairs.

As shown in Figs. 3-10, our proposed scheme outperforms the scheme in [7].

## VI. Conclusions

We proposed a scheme for increasing the total capacity of fully loaded 4G cellular networks while minimizing the total uplink transmit power when underlay D2D links are utilized. We maintain the required QoS in terms of SINR for all users, and show that energy efficiency is increased. We solved the optimization problem by dividing it into two sub problems and solving each sub problem individually. Simulation results demonstrate the improvements in using our proposed scheme.

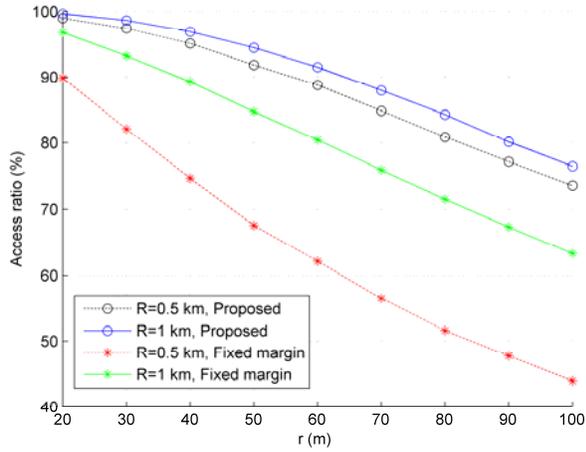

Fig. 3. D2D access ratio for different cell and D2D cluster radii, where $N = 20$ and $M = 10$.

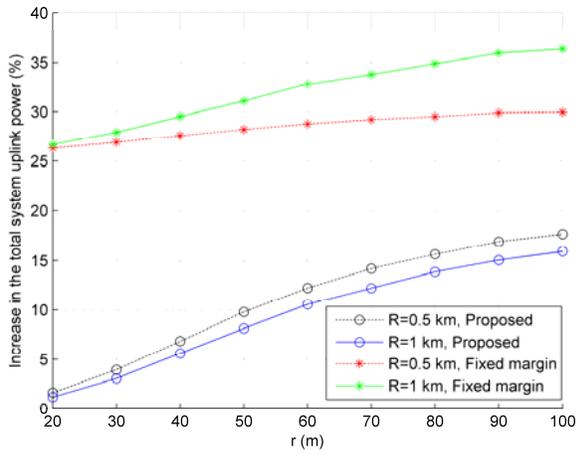

Fig. 4. Increase in the total system uplink power for different cell and D2D cluster radii, where $N = 20$ and $M = 10$.

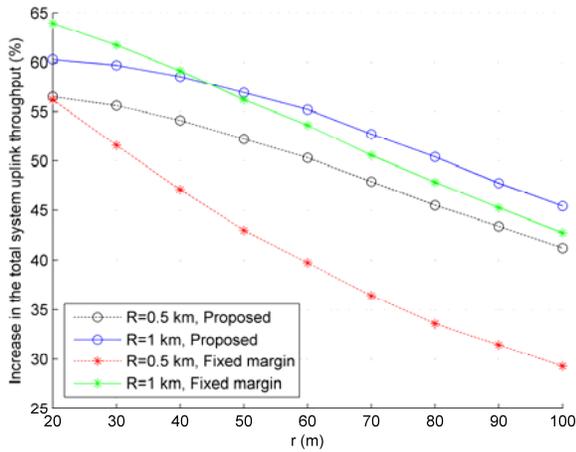

Fig. 5. Increase in the total system uplink throughput for different cell and D2D cluster radii, where $N = 20$ and $M = 10$.

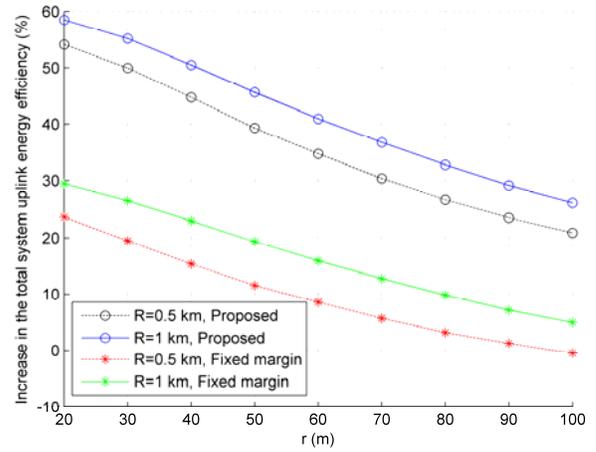

Fig. 6. Increase in the total system uplink energy efficiency for different cell and D2D cluster radii, where $N = 20$ and $M = 10$.

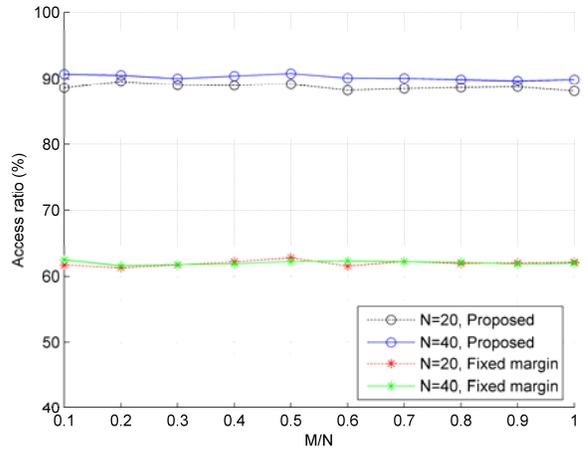

Fig. 7. D2D access ratio for different number of active CUs and D2D pairs, where $R = 0.5$ km and $r = 60$ m.

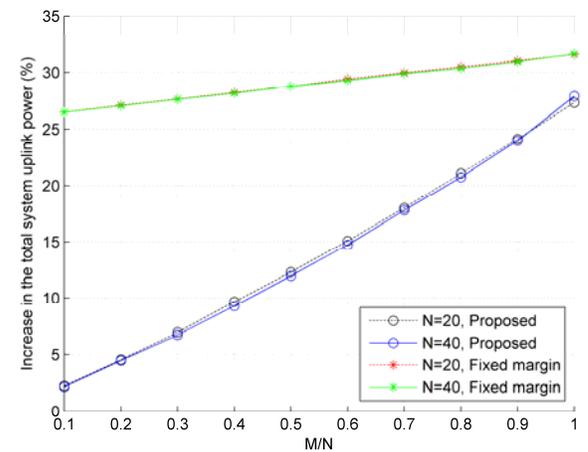

Fig. 8. Increase in the total system uplink power for different number of active CUs and D2D pairs, where $R = 0.5$ km and $r = 60$ m.

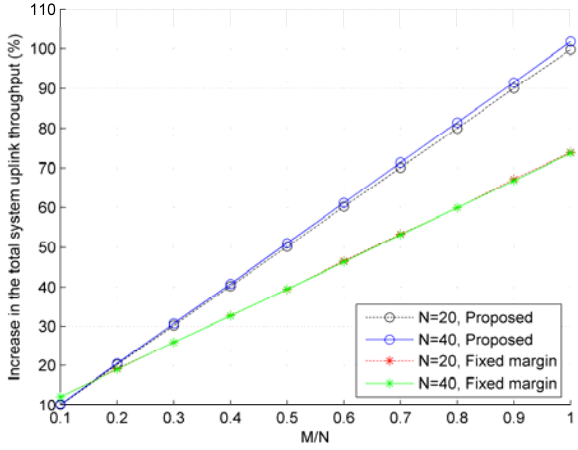

Fig. 9. Increase in the total system uplink throughput for different number of active CUs and D2D pairs, where $R = 0.5$ km and $r = 60$ m.

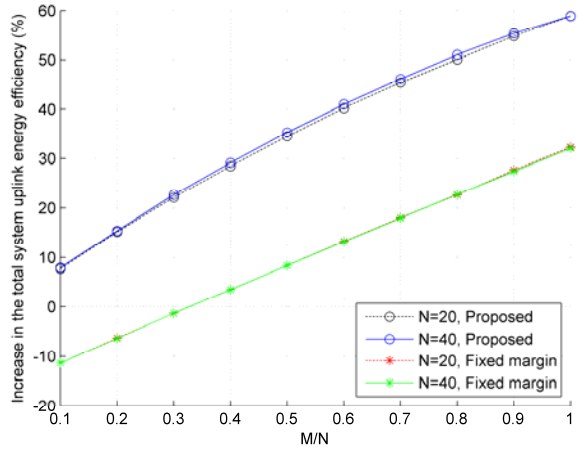

Fig. 10. Increase in the total system uplink energy efficency for different number of active CUs and D2D pairs, where $R = 0.5$ km and $r = 60$ m.

References

[1] OECD, *Broadband and the Economy*, Ministerial background Report, OECD Ministerial Meeting on the Future of the Internet Economy, Seoul, Korea, 17-18 June 2008.

[2] Cisco, *Cisco Visual Networking Index: Global mobile data traffic forecast update, 2013–2018*. Available [Online]: http://www.cisco.com/c/en/us/solutions/collateral/service-provider/visual-networking-index-vni/white_paper_c11-520862.html

[3] K. Doppler, M. Rinne, C. Wijting, C. B. Ribeiro, and K. Hugl, "Device-to-device communication as an underlay to LTE-advanced networks," *IEEE Communications Magazine*, vol. 47, no. 12, pp. 42-49, December 2009.

[4] M. J. Yang, S. Y. Lim, H. J. Park, and N. H. Park, "Solving the data overload: device-to-device bearer control architecture for cellular data offloading," *IEEE Vehicular Technology Magazine*, vol. 8, no. 1, pp. 31-39, March 2013.

[5] G. Fodor, E. Dahlman, G. Mildh, S. Parkvall, N. Reider, G. Miklos, and Z. Turanyi, "Design aspects of network assisted device-to-device communications," *IEEE Communications Magazine*, vol. 50, no. 3, pp. 170-177, March 2012.

[6] C.-H. Yu, K. Doppler, C. B. Ribeiro, and O. Tirkkonen, "Resource sharing optimization for device-to-device communication underlaying cellular networks," *IEEE Transactions on Wireless Communications*, vol. 10, no. 8, pp. 2752–2763, August 2011.

[7] B. Kaufman, J. Lilleberg, and B. Aazhang, "Spectrum sharing scheme between cellular users and ad-hoc device-to-device users," *IEEE Transactions on Wireless Communications*, vol. 12, no. 3, pp. 1038-1049, March 2013.

[8] P. Phunchongharn, E. Hossain, and D. I. Kim, "Resource allocation for device-to-device communications underlaying LTE-advanced networks," *IEEE Wireless Communications*, vol. 20, no. 4, pp. 91-100, August 2013.

[9] D. Feng, L. Lu, Y. Yuan-Wu, G. Y. Li, G. Feng, and S. Li, "Device-to-device communications underlaying cellular networks," *IEEE Transactions on Communications*, vol. 61, no. 8, pp. 3541-3551, August 2013.

[10] Z. Hasan, H. Boostanimehr and V. K. Bhargava, "Green Cellular Networks: A Survey, Some Research Issues and Challenges," *IEEE Communications Surveys & Tutorials,* vol. 13, no. 4, pp. 524-540, 2011.

[11] X. Xiao, X. Tao, and J. Lu, "A QoS-aware power optimization scheme in OFDMA systems with integrated device-to-device (D2D) communications," *Proceedings of IEEE VTC-Fall*, Sept. 2011, pp. 1–5.

[12] U. Varshney, "4G Wireless Networks*," IEEE IT Professional*, vol. 14, no. 5, pp. 34-39, September-October 2012.

[13] M. Bazaraa, J. Jarvis and H. Sherali, *Linear Programming and Network Flows*, Wiley, 2010.

[14] K. G. Murty, *Network Programming*, Prentice Hall, 1992.